\documentstyle[preprint,prl,aps]{revtex}

\begin{document}
\title{Comment on ``New Experimental Limit on the Electric Dipole Moment of the
Neutron"}
\author{S.K. Lamoreaux$^1$ and R. Golub$^2$}
\address{$^1$University of California, \\
Los Alamos National Laboratory,\\
Physics Division P-23, M.S. H803, Los Alamos, New Mexico 87545\\
$^2$ Hahn-Meitner Institut, Glienicker Str. 100, D-14109 Berlin,\\
Germany}
\date{\today}
\maketitle

\begin{abstract}
A new limit for the neutron electric dipole moment has been recently
reported. This new limit is obtained by combining the result from a previous
experiment with the result from a more recent experiment that has much worse
statistical accuracy. 
We show that the old result has a systematic error
possibly four times greater than the new limit, and under the circumstances,
averaging of the old and new results is statistically invalid. The conclusion
is that it would be more appropriate to quote two independent but mutually
supportive limits as obtained from each experiment separately.

\end{abstract}


\tightenlines

Recently, a new experimental limit for the neutron permanent electric dipole
moment (EDM) has been reported \cite{new} with a substantial increase in
sensitivity as compared to the previous result \cite{old}. The increase in
sensitivity results from combining the old experimental result with new data
that has considerably worse statistical accuracy. As we discuss below, the
averaging of the old and new results is not justified in this situation and
one would better quote two independent results that are mutually consistent
with and supportive of each other. It makes no sense to average the results
because there are unexplained systematic errors in the old data that exceed
the new 95\% confidence limit given in \cite{new} by a factor of at least
two.

The new quoted 90\% confidence limit of $|d_n|<6.3\times 10^{-26}\ e{\rm cm}$
was obtained by combining the result of the new (1999) experiment 
\[
d_n=1.9\pm 5.4\times 10^{-26}\ e{\rm cm};\ \ \ \chi^2/\nu=0.4 
\]
(see Fig. 1, where $\nu=9$) with the previous (1990) result, 
\begin{equation}
-3\pm 5\times 10^{-26}\ e{\rm cm}
\end{equation}
for which the error is dominated by systematic uncertainty in the data. The
95\% confidence limit for the 1990 data is given as 
\begin{equation}  \label{95}
|d_n|<12\times 10^{-26}\ e {\rm cm}.
\end{equation}

Without attempting to correct for systematic effects, the average of the
1990 data shown in Fig. 1 is 
\begin{equation}
d_{n}=-1.9\pm 2.2\times 10^{-26}\ e{\rm cm};\ \ \ \chi ^{2}/\nu =3.2
\end{equation}
and $\nu =14$. Unfortunately, there is no clear way to use the discrete Rb
magnetometers employed in the 1990 version of the experiment to correct for
the magnetic systematic effect that is evidenced through the large $\chi
^{2}/\nu$ and bimodal character of the data; simply subtracting the average
of the Rb magnetometer signals gives the result 
\begin{equation}
d_{n}=-(3.4\pm 2.6)\times 10^{-26}\ e{\rm cm},\ \ \ \ \chi ^{2}/\nu =2.2;
\label{goo}
\end{equation}
where the reduction in $\chi ^{2}$ is due to the increase in the effective
standard deviation due to the magnetometer noise, not a decrease in the
scatter of the data, as is stated in \cite{old}.

The discussion associated with Eq. (5) of \cite{new} (also labelled here as
Eq. (5)), 
\begin{equation}
d_{n}=-3.4\pm 3.9\ \ {\rm stat}\ \ \pm 3.1\ \ {\rm syst}\ \ \times 10^{-26}\
e{\rm cm}  \label{eq5}
\end{equation}
does not accurately describe the source of errors in the final quoted
result. In particular, the statistical error in Eq. (5) is obtained by
multiplying the uncertainty in Eq. (4) by the square root of reduced $\chi
^{2}$, $2.6\times \sqrt{2.2}=3.9$. Therefore, the statistical error given in
Eq. (5) has a substantial systematic contribution. Because there is no
physical basis for subtracting the average Rb magnetometer signal from the
neutron EDM result (as we will describe), if instead we take the intrinsic
neutron EDM statistical error as that given in Eq. (3), Eq. (5) implies a
systematic error of $\sqrt{3.9^{2}-2.2^{2}}=3.2\ (\times 10^{-26}\ e{\rm cm}%
) $; this should be added in quadrature to the systematic error given in Eq.
(5) yielding an overall systematic uncertainty of at least 
\begin{equation}
4.8\times 10^{-26}\ e{\rm {cm}}  \label{sys}
\end{equation}
and represents a guess at how well the average systematic fluctuations
averaged to zero for the entire 1990 data set. This systematic error, when
combined with the intrinsic statistical error, yields the combined
uncertainty given in Eq. (1).

In fact, the final systematic uncertainty was chosen so that the 95\%
confidence limit would encompass the obvious excess scatter in the 1990
data, reflected by the large $\chi^2/\nu=3.2$. The final 95\% confidence
limit was chosen to reflect this deviation in the data because it was
accepted that there is no reason to assume the random systematic
fluctuations average to zero; Eq. (1) above might very well be optimistic.
These fluctuations are correlated with disassembly and reassembly of the
apparatus, and therefore are possibly the result of subtle changes in, for
example, high-voltage leakage current paths within the magnetic shields.

We would like to take a different approach to combining the 1990 and 1999
data groups; this approach gives a more accurate estimate of a possible
systematic error in the final combined number. As shown in Fig. 1, the 1990
data is better described by a bimodal distribution. Curve A is a $1/\sigma
^{2}$ bimodal weighted fit to two Gaussian probability distributions for the
1990 data alone. Curve B shows the bimodal probability distribution for the
1990 and 1999 combined data sets, while C is the combined data set when the
standard errors of the 1990 data are multiplied by $\sqrt{3.2}$. It is
evident that the distribution for the combined data set, for both B and C,
is still strongly bimodal and the use of standard statistical techniques as
in \cite{new} is not justified.

This is demonstrated in Fig. 2 which gives the 68.3\% and 95\% confidence
regions ($\Delta \chi ^{2}=2.3$ and $\Delta \chi ^{2}=6.17$, respectively,
for two degrees of freedom \cite{numrec}) for the two means determined in
the fit to the bimodal distribution (a total of six parameters in the fit).
Of particular interest is Fig. 2 A; $\chi ^{2}_{19}$ for the optimum fit to
two Gaussian distribution (each Gaussian has three fit parameters so there
are $25-6=19$ degrees of freedom) is 8.3 as compared to $\chi ^{2}_{22}=48$
for a fit to a single Gaussian distribution. The reduced $\chi^2$ for the
bimodal fit is $\chi ^{2}_{19}/19 =0.44$. However, the statistical
significance is best illustrated by use of the $F$-test \cite{bev}. Adding
the second Gaussian distribution is equivalent to adding three fit
parameters, and 
\begin{equation}
F={\frac{\left[ \chi ^{2}_{22}-\chi ^{2}_{19}\right] /3}{\chi ^{2}_{19}/19}}=%
{\frac{\left( 48-8.3\right) /3}{(.44\rightarrow 1)}}=13.2
\end{equation}
(the division by the new reduced $\chi ^{2}/\nu$ is valid only when it is
greater than 1). The statistical confidence of the new bimodal description
of the combined data can be calculated (see Appendix C-5 of \cite{bev}); the
probability of such a large $F$ being due to statistical fluctuations is
about 3\%, {\em i.e. } the statistical confidence is 97\%. There is no
justification for artificially increasing the errors of the 1990 data by $%
\sqrt{3.2}$ because the final combined data set without doing so as
described by the bimodal distribution has a reduced $\chi ^{2}/\nu =0.44$.
We again point out that even if the 1990 errors are increased by $\sqrt{3.2}$
the full data set is still bimodal, as shown in Fig. 1 C and Fig. 2 B.

The bimodal means are of opposite sign, but of remarkably close magnitudes: 
\begin{equation}
d_{n1}=12.1\pm 5.0;\ \ d_{n2}=-11.7\pm 4.5\ \times 10^{-26}\ e{\rm cm}
\end{equation}
and represent the magnitude of the random systematic in the data. In Eq.
(1), there is an implied faith that the systematic errors in the data
average to zero, and therefore Eq. (1) must be treated with a certain
caution. The Particle Data Group does indeed recommend scaling of the data
by $\sqrt{\chi _{\nu }^{2}}$ as done in [1]. However they justify this by
the remark that this approach has the property that if there are two values
with comparable errors separated by much more than their stated errors the
error on the mean is increased so that it is approximately half the interval
between the two discrepant values. Applying this criterion to the present
case and taking the bimodal means for the 1990 data as two discrepant values
shows a systematic error of 
\begin{equation}
{\frac{d_{n1}-d_{n2}}{2}}=12\times 10^{-26}\ e{\rm cm}
\end{equation}
which is rather larger than the weighted error when scaled by $\sqrt{3.2}$, 
{\em i.e.} a factor of two greater than Eq. (6). This implies a 95\%
confidence limit of about $20\times 10^{-26}\ e{\rm cm}$, about a factor of
two greater than Eq. (2). In addition, both $\chi ^{2}$ plots in Fig. 2
indicate a net 95\% confidence limit of about $20\times 10^{-26}\ e{\rm {cm}}
$. Under the circumstances, the 95\% confidence limit given in \cite{old},
Eq. (2) above, requires additional justification.

Furthermore, other analysis techniques support this magnitude of random
systematic for the 1990 data. For example, when only data where the
magnetometer signals were statistically zero was included in the average, a
greater than two-sigma hint of a non-zero value the neutron EDM was obtained 
\cite{aix}, 
\begin{equation}
d_{n}=-(17\pm 6)\times 10^{-26}\ e{\rm {cm}.}
\end{equation}
without correcting for the magnetometers. There is no statistical or
physical basis for subtracting the average magnetometer signal, as it has
been proven by use of correlation techniques \cite{edm,cp} that even when
the average magnetometer signal is zero, the individual magnetometers are
still correlated with the electric field, and with the neutron EDM signal.
This implies that the magnitude, and possibly sign, of the correlated field
measured by the Rb magnetometers is different among the three magnetometers
and the neutron storage cell.

A better correction for a possible systematic effect was obtained by
determining the correlation coefficient between each magnetometer and the
neutron EDM signal. Subtracting the correlated Rb magnetometer signal made
the neutron data internally more consistent; in this analysis, before
subtracting the correlated signal, $d_{n}=-(2.8\pm 3.0)\times 10^{-26}\ e%
{\rm {cm}}$, $\chi ^{2}=3.8$, and after subtraction, $d_{n}=-(5.0\pm
3.8)\times 10^{-26}\ e{\rm {cm}}$, $\chi ^{2}=1.4$. The decrease in $\chi
^{2}$ in this case results from a reduction of scatter rather than an
increase in the effective per point standard deviation due to magnetometer
noise.

Agreement between the various analyses was used to estimate the residual
random systematic in the EDM signal that did not average away, and in
particular, was chosen so that the 95\% confidence limit of the combined
statistical and systematic error encompassed the range of systematic
fluctuation, as mentioned above. Although the correlation technique is
compelling, it is not proof that the systematic contribution has been fully
accounted for. The implication is that the average field seen by the
magnetometers does not necessarily represent the field seen by the neutrons,
and was the motivation for construction of the $^{199}$Hg co-magnetometer 
\cite{skl}.

To our knowledge, the correlation techniques were not applied to the new
data, and therefore the 1999 data cannot be used to infer anything about the
source of the random systematic in the 1990 data. In particular, we disagree
with the statements associated with Eq. (5) of \cite{new}; the sources of
the scatter in the previous data are still unknown, and experimental
evidence suggests they are associated with the magnetic fluctuations
generated with the application of high voltage; this is the systematic of
principal concern. Although the statements in \cite{new} regarding magnetic
fluctuations are technically correct, the ``other, unknown, systematics''
suggested in \cite{new}, if they even exist, are likely irrelevant to any
discussion. Again, the important point is whether the magnetic fluctuations
that contaminated the data are associated with the high voltage or with
other noise sources.

The new data sheds no light on the previous result, particularly when one
considers the vast array of changes with the new configuration of the
apparatus. For example, the much larger ultracold neutron storage vessel in
the new experiment means that a discrete magnetometer placed within the
shield would be relatively farther from the geometrical center of the
neutron storage region, so the relative effect of a leakage current magnetic
field could be quite different between the previous and present
configurations of the apparatus; although Rb magnetometers were included in
the new version of the apparatus, no presentation or discussion of
experimental data from them is given in \cite{new}. Of more critical
concern, in \cite{new} the apparent $^{199}$Hg EDM in the co-magnetometer
data is not given, and is crucial toward understanding possible systematic
effects. The graph given in Fig. 3 of \cite{new} might seem compelling, but
tells us nothing in regard to subtle effects with application of high
voltage; the fluctuations shown in Fig. 3 of \cite{new} are likely primarily
due to external influences. In regard to the old data, it was accepted that
external magnetic influences on the neutron EDM are well-represented by the
average Rb magnetometer signal, while local fields, particularly those
associated with the high voltage supply or leakage current, are not.

Under the circumstances, the result for the 1990 data, Eq. (5), should be
treated with a certain caution. The final uncertainty in this result was
chosen so that the 95\% confidence limit would encompass the magnitude of
the random systematic, $\approx 12\times 10^{-26}\ e{\rm {cm}}$. Nothing
associated with the new data would allow one to assume that the rather large
random systematic averaged to zero in the previous data. It therefore is
imprudent to average the old and new results. Hopefully, in the
not-too-distant future, a co-magnetometer neutron EDM experiment will be
possible with ultracold neutron density equal to or much greater than that
obtained in connection with the 1990 version of the experiment. Only then
will a truly new and improved limit for the neutron EDM, without concerns of
systematic contamination of up to $12\times 10^{-26}\ e{\rm {cm}}$ or
greater, be possible.

We thank Michael Romalis for helpful discussions.

\begin{figure}
\caption{Previous and new neutron EDM data averaged over
subsequent several-week reactor operation cycles.  The horizontal
axis is arbitrary and labels the chronological order of the
data subsets.  Particulary for the 1990 data, the apparatus
was dis- and re-assembled between reactor cycles. A: Fit to a 
bimodal distribution for the 1990 data.  B: Fit to a bimodal
distribution for the combined 1990 and 1999 data.  C:  Fit to
a bimodal distribution with the 1990 data errors
multiplied by $\sqrt{3.2}$.}
\end{figure}

\begin{figure}
\caption{68.3\% and 95\% confidence regions for the two means in
the bimodal fits.  A: Combined 1990 and 1999 data.  B. Combined
data, 1990 data error multiplied by $\sqrt{3.2}$.}
\end{figure}

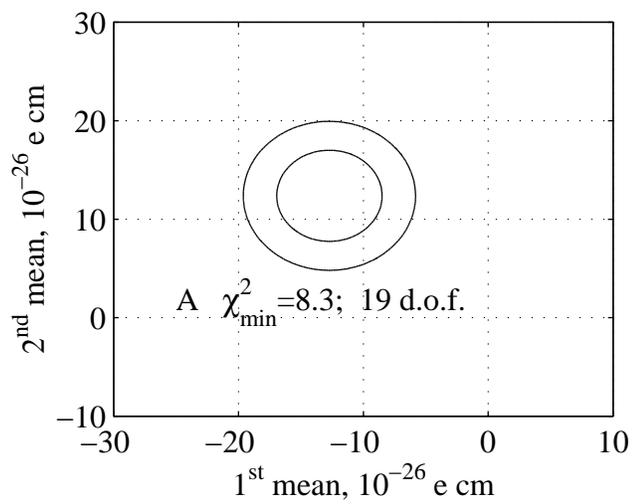
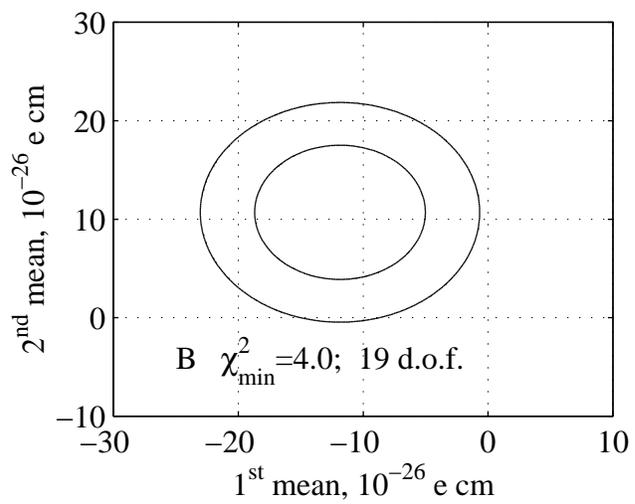
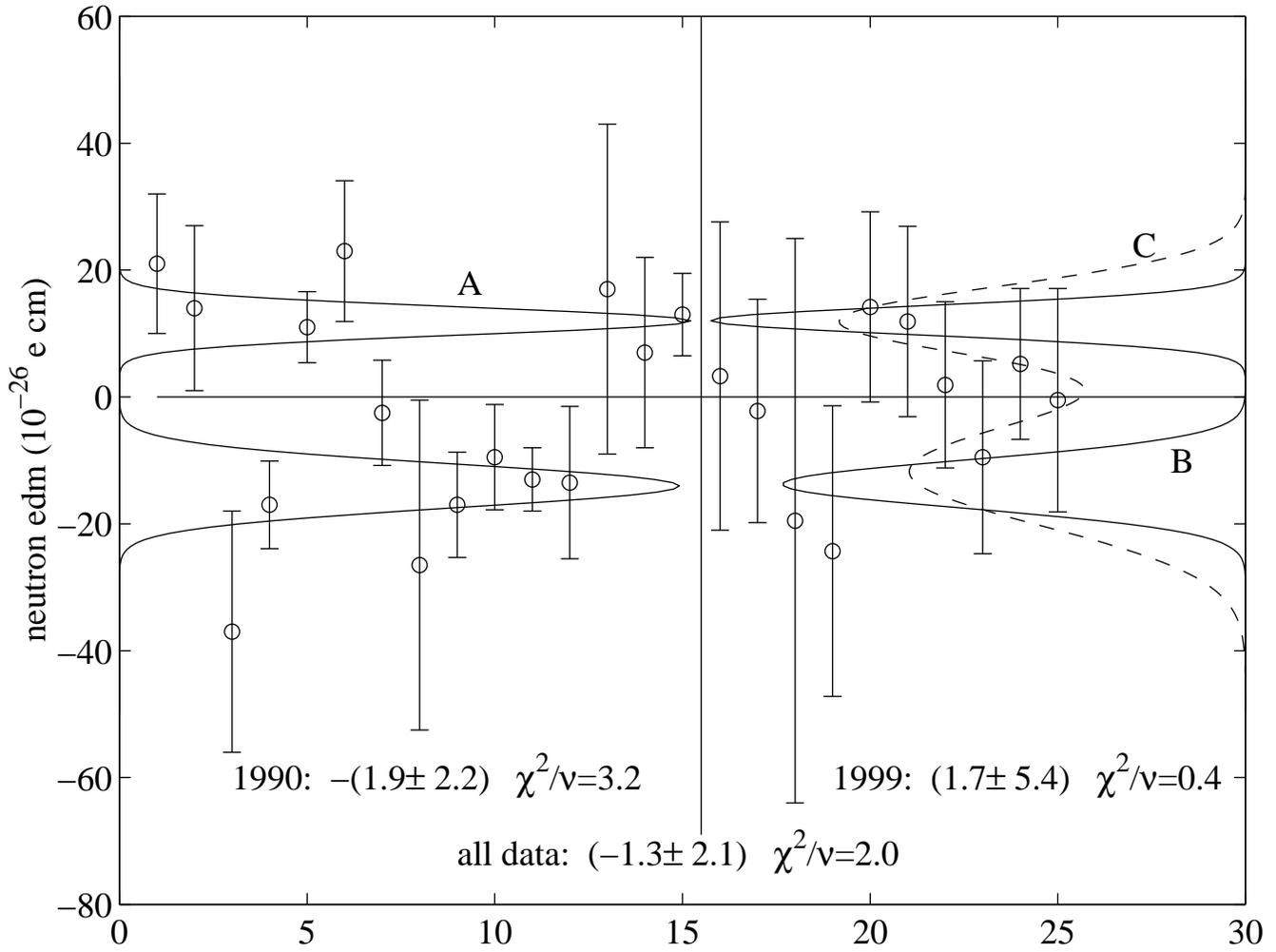
\end{document}